\documentclass[showkeys,preprint,twocolumn,11pt,times,superscriptaddress,pre]{revtex4-2}
\usepackage{stix}
\usepackage[T1]{fontenc}
\usepackage[english]{babel}
\usepackage{graphicx}
\usepackage[outdir=./]{epstopdf}
\usepackage{bm} 
\usepackage[colorlinks=true,
            linkcolor=blue,
            urlcolor=blue,
            citecolor=blue]{hyperref}
\usepackage[mathlines]{lineno} 
\usepackage{amsmath}
\usepackage{physics}
\usepackage{booktabs}
\usepackage{lipsum}
\usepackage{geometry}
\begin{document}
    \title{Ratchet current and scaling properties in a nontwist mapping}

    \author{Matheus Rolim Sales}
    \email{matheusrolim95@gmail.com}
    \affiliation{Departamento de Física, Universidade Estadual Paulista (UNESP), 13506-900, Rio Claro, SP, Brasil}
    \author{Daniel Borin}
    \affiliation{Departamento de Física, Universidade Estadual Paulista (UNESP), 13506-900, Rio Claro, SP, Brasil}
    \author{Leonardo Costa de Souza}
    \affiliation{Instituto de Física, Universidade de São Paulo, 05315-970, São Paulo, SP, Brasil}
    \author{José Danilo Szezech Jr.}
    \affiliation{Programa de Pós-Graduação em Ciências, Universidade Estadual de Ponta Grossa, 84030-900, Ponta Grossa, PR, Brasil}
    \affiliation{Departamento de Matemática e Estatística, Universidade Estadual de Ponta Grossa, 84030-900, Ponta Grossa, PR, Brasil}
    \author{Ricardo Luiz Viana}
    \affiliation{Departamento de Física, Universidade Federal do Paraná, 81531-990, Curitiba, PR, Brasil}
    \author{Iberê Luiz Caldas}
    \affiliation{Instituto de Física, Universidade de São Paulo, 05315-970, São Paulo, SP, Brasil}
    \author{Edson Denis Leonel}
    \affiliation{Departamento de Física, Universidade Estadual Paulista (UNESP), 13506-900, Rio Claro, SP, Brasil}
    \date{\today}

    \begin{abstract}
        We investigate the transport of particles in the chaotic component of phase space for a two-dimensional, area-preserving nontwist map. The survival probability for particles within the chaotic sea is described by an exponential decay for regions in phase space predominantly chaotic and it is scaling invariant in this case. Alternatively, when considering mixed chaotic and regular regions, there is a deviation from the exponential decay, characterized by a power law tail for long times, a signature of the stickiness effect. Furthermore, due to the asymmetry of the chaotic component of phase space with respect to the line $I = 0$, there is an unbalanced stickiness which generates a ratchet current in phase space. Finally, we perform a phenomenological description of the diffusion of chaotic particles by identifying three scaling hypotheses, and obtaining the critical exponents via extensive numerical simulations.
    \end{abstract}

    \maketitle

    \section{Introduction}
    \label{sec:introduction}

    The phase space of a two-dimensional integrable Hamiltonian system is composed of periodic and quasiperiodic invariant tori. When a weak perturbation is introduced into such a system, according to the Kolmogorov-Arnold-Moser (KAM) theorem~\cite{lichtenberg2013regular}, the sufficiently irrational tori survive the perturbation (KAM tori), while the rational ones are destroyed. Near the original position of the destroyed rational tori, emerges a set of elliptical and hyperbolic fixed points, as outlined by the Poincaré-Birkhoff theorem~\cite{lichtenberg2013regular}. The chaotic motion appears in the vicinity of the hyperbolic fixed points due to their unstable nature, while the elliptical points are the centers of the regular regions, hereafter named stability islands. The coexistence of chaotic and regular regions in two-dimensional quasi-integrable Hamiltonian system makes its phase space neither integrable nor uniformly hyperbolic. The phase space is divided into distinct and unconnected domains, where the chaotic orbits fill densely the available region in phase space and the stability islands consist of periodic and quasiperiodic orbits that lie on invariant tori. Furthermore, an orbit initially in the chaotic sea will never enter any island, and the periodic and quasi-periodic orbits will never reach the chaotic sea~\cite{lichtenberg2013regular, Mackay1984b}.
    
    For increasing perturbation strength, the KAM tori are also destroyed and their remnants form a Cantor set known as cantori~\cite{Mackay1984a,Mackay1984b,Efthymiopoulos1997}. The role of the KAM tori and the cantori in the transport of particles in phase space differs fundamentally. While the KAM tori are full barriers to the transport in phase space, the cantori act as partial barriers, allowing particles to pass through them. When a particle crosses a cantorus, it might stay trapped in the region bounded by the cantorus for a long, but finite, period of time, during which it behaves similarly as a quasiperiodic orbit, until it escapes to the chaotic sea. This intermittence in the dynamics of a chaotic orbit is the phenomenon of stickiness ~\cite{Contopoulos1971, KARNEY1983360, MEISS1983375, CHIRIKOV1984395, Efthymiopoulos1997, Contopoulos2008, Cristadoro2008, Contopoulos2010,Zaslavsky2002}. The structure of stability islands embedded in the chaotic sea and cantori organize itself in a hierarchical structure of islands-around-islands, where the larger islands are surrounded by smaller islands, which are in turn surrounded by even smaller islands and so on for increasingly smaller scales~\cite{Meiss1985,Meiss1986}. In this way, during the time a chaotic orbit is trapped in a region bounded by a cantorus, it might cross inner cantori for arbitrarily small scales and thus stay trapped for longer times. This long times affects the statistical properties of the transport of particles in phase space~\cite{Zaslavski1972, Zaslavsky1997, ZASLAVSKY2002461, zaslavsky2005hamiltonian}, as well as the recurrence time statistics~\cite{Afraimovich1997, Altmann2005, Altmann2006,Venegeroles2009, Abud2013, Lozej2020}, the survival probability~\cite{Lai1992, Cristadoro2008,Altmann2009, Avetisov2010, DETTMANN2012403, LEONEL20121669,Livorati2012, Livorati2014,Mendez-Bermudez_2015,deFaria2016, LIVORATI2018225,Borin2023}, and the decay of correlations~\cite{KARNEY1983360,MEISS1983375,Vivaldi1983, CHIRIKOV1984395, Lozej2018}.

    About two decades ago, a new feature was observed in the transport of particles in the chaotic component of phase space in Hamiltonian systems: the existence of a preferential direction for the transport without an external bias, the so-called Hamiltonian ratchet~\cite{Dittrich2000, Denisov2001,Schanz2001, Denisov2002,Monteiro2002,Cheon2003,Gong2004}. The ratchet effect is defined by a directed current, or ratchet current, which is a preferential direction for the transport of particles in phase space without an external bias. This phenomenon has applications on a variety for physical systems such as Josephson junctions~\cite{josephjunctions,Zapata1996}, Brownian~\cite{Reimann2002} and moleculars~\cite{RevModPhys.69.1269} motors, cold atom systems~\cite{coldatoms1,coldatoms2,coldatoms3,coldatoms4}, and eletronic transport in superlattice~\cite{bass1997kinetic,Schelin2010}, to cite a few. Inital studies focused on the role of the external noise~\cite{Bartussek1994, Luczka1997, Reimann2002}, which was later replaced by a determistic chaotic dynamics with inertia terms in the equations of motion~\cite{Jung1996, Flach2000, Mateos2000, Porto2000}. Nonetheless, even purely Hamiltonian dynamics with mixed~\cite{Dittrich2000, Denisov2001,Denisov2002,Schanz2001,Cheon2003,Gong2004} or completly chaotic~\cite{Monteiro2002} phase space can generate ratchet current. It was demonstrated that the ratchet effect is a consequence of spatial and/or temporal symmetry breaking of the system~\cite{Denisov2001,Denisov2002,Gong2004,Wang2007}. Furthermore, Mugnaine \textit{et al.}~\cite{Mugnaine2020} demonstrated that due to a symmetry breaking in the extended standard nontwist mapping~\cite{Portela2007, Wurm2013}, the twin island scenario no longer exists, implyng a ratchet current in phase space due to the emergence of an unbalanced stickiness in different regions in phase space.

    The anomalous transport of particles in magnetized plasmas is heavily impacted by the $\vb{E}\times\vb{B}$ drift motion, caused by fluctuations in the plasma \cite{balescu2005aspects}. In the case of passive particle transport, where the particles do not alter the electric field, the problem is described by a Hamiltonian system \cite{Horton_1985}. Furthermore, given the divergence-free nature of magnetic fields, the field lines can be described by a two-dimensional area-preserving mapping, with respect to the surface of section of the torus at a fixed toroidal angle \cite{10.1063/5.0170345}. In this paper, we study transport properties and diffusion for a nontwist mapping, introduced in the context of $\vb{E} \times \vb{B}$ drift in toroidal plasmas~\cite{horton1998drift,10.1063/5.0147679,PhysRevE.109.015202}, in the transition from integrability (zero perturbation) to non-integrability (small perturbation) when the phase space is bounded.  In the context of magnetic confined plasmas, as in Tokamak devices~\cite{plasmas,weiland2019drift}, the use of non-monotonic profiles for physical quantities is related to nontwist effects like the existence of a shearless curve that acts as a robust barrier preventing the chaotic orbits from escaping the plasma \cite{CALDAS20122021}. Nontwist Hamiltonian systems are characterized by complex dynamics, including the formation of transport barriers and intricate island structures. Understanding these chaotic transport processes is essential for predicting and controlling plasma behavior.
    
    Initially, we analyze the survival probability by introducing two exits symmetrically apart from the line $I = 0$ and show that as long as there are no stability islands within the survival regions, the survival probability follows an exponential decay, and the decay rate scales as a power law with the limits of the survival regions. Furthermore, the survival probability exhibits scaling invariance with respect to the limits of the survival regions. Then, we consider two different ensembles of initial conditions, both with $\langle I \rangle \approx 0$ initially. The first ensemble is uniformly distributed along the whole survival region and we calculate the escape times and the escape basins for different survival region limits. We find that the measure of the bottom basin increases with the survival region area due to the inclusion of stability islands within this region. This ensemble, however, might be biased due to the distribution of initial conditions in phase space. To avoid this potential bias, we consider a second ensemble randomly distributed in a small region around the line $I = 0$ and calculate the fraction of particles that escape from the top and bottom exits. We find, once again, that as we change the size of the survival region and islands start to appear, there is a tendency for the particles to escape through the bottom exit. Also, we observe a nonzero space average of the action, $\langle I \rangle$, characteristic of the ratchet effect.
    
    As for the mechanism behind such a phenomenon, we calculate the distribution of recurrence times considering the upper ($I > 0$) and lower ($I < 0$) regions of phase as our recurrence regions. We find that the cumulative distribution of recurrence times for these regions are different for long times, indicating that the chaotic orbits sticky unevenly in the upper and lower regions, \textit{i.e.}, there is an unbalanced stickiness in phase space. The first and higher moments of the distribution of recurrence times are also different, consolidating that the two distributions are, in fact, different. Lastly, we investigate the scaling properties of diffusion in phase space. We choose as our observable the square root of the averaged squared action, $I_{\mathrm{rms}}$, and we obtain the critical exponents that describe the behavior of $I_{\mathrm{rms}}$. These quantities are described in terms of three scaling hypotheses, leading to a robust analysis of the scaling invariance observed for our observable.

    This paper is organized as follows. In Section~\ref{sec:mapping} we describe the mapping under study and present some of its properties. In Section~\ref{sec:survprob} we study the transport properties of the chaotic component of phase space. We begin by calculating the survival probability for different survival regions and in the sequence, we investigate whether the system exhibits unbiased transport by considering two different ensembles of initial conditions with zero initial average action. In Section~\ref{sec:Irms} we present a phenomenological description of diffusion in the chaotic component of phase space. We obtain the critical exponents and compare our results with similar results in the literature. Section~\ref{sec:finalremarks} contains our final remarks.

    \section{The model and its properties}
    \label{sec:mapping}

    \begin{figure*}[tb]
        \centering
        \includegraphics[width=\linewidth]{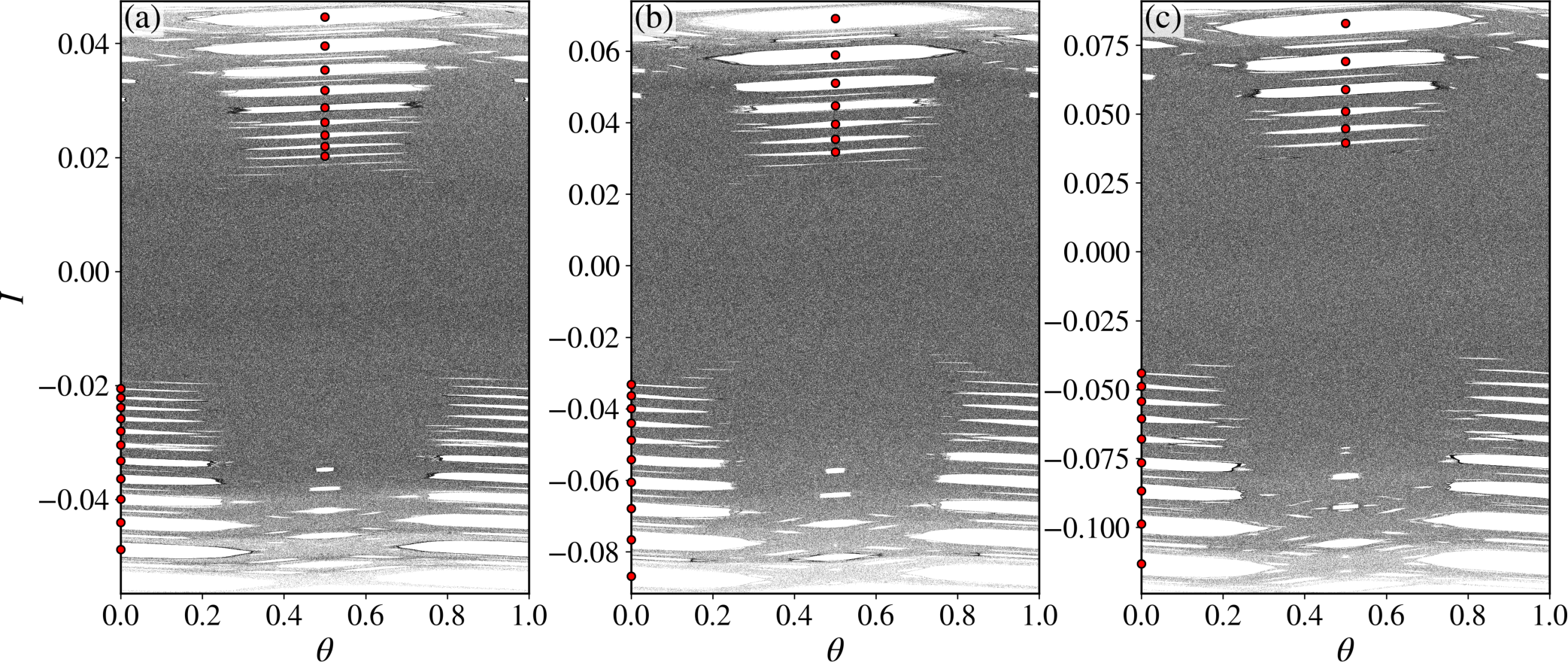}
        \caption{The phase space of the mapping Eq.~\eqref{eq:mapa} for 375 randomly chosen initial conditions within the chaotic sea with (a) $\varepsilon = 1.0\times10^{-3}$, (b) $\varepsilon = 2.0\times10^{-3}$, and (c) $\varepsilon = 3.0\times10^{-3}$. The red dots correspond to elliptic points at the center of the period-1 islands.}
        \label{fig:phasespace}
    \end{figure*}

    The Hamiltonian function of an autonomous two degrees of freedom system is often written as~\cite{lichtenberg2013regular}
    \begin{equation}
        \label{eq:genham}
        H(I_1, I_2, \theta_1, \theta_2) = H_0(I_1, I_2) + \varepsilon H_1(I_1, I_2, \theta_1, \theta_2),
    \end{equation}
    where $(I_i, \theta_i)$ are the canonical action-angle variables, $H_0$ is the integrable term and $\varepsilon H_1$ is a perturbation to the integrable system, with $\varepsilon$ controlling the transition from integrability ($\varepsilon = 0$) to non-integrability ($\varepsilon > 0$). The solution of Eq.~\eqref{eq:genham} is a four-dimensional flow. However, since $\pdv*{H}{t} = 0$, the Hamiltonian equals the total mechanical energy of the system, $H = E = T + V$, and it is a constant of motion. This allows us to eliminate one of the variables, \textit{e.g.} $I_2$, from $H$, making it possible to write $H = H(I_1, \theta_1, \theta_2, E)$. Therefore, orbits with energy $E$ are restricted to lie on a three-dimensional energy surface on the four-dimensional phase space. It is possible to decrease the dimension even further by considering an appropriate Poincaré section. We choose the plane $\theta_1 \times I_1$ with $\theta_2$ constant as our Poincaré section, resulting in a two-dimensional mapping described by the following equations:
    \begin{equation}
        \label{eq:genmapping}
        \begin{aligned}
            I_{n + 1} &= I_n + \varepsilon h(\theta_n, I_{n + 1}),\\
            \theta_{n + 1} &= \theta_n + K(I_{n + 1}) + \varepsilon p(\theta_n, I_{n + 1}) \mod2\pi,
        \end{aligned}
    \end{equation}
    where $h(\theta_n, I_{n + 1})$, $K(I_{n + 1})$, and $p(\theta_n, I_{n + 1})$ are nonlinear functions of their arguments and this mapping relates the $(n + 1)$th intersection and the previous $n$th intersection with the Poincaré section. The mapping in Eq.~\eqref{eq:genmapping} is area-preserving if the nonlinear functions satisfy the following condition
    \begin{equation}
        \pdv{p(\theta_n, I_{n + 1})}{\theta_n} + \pdv{h(\theta_n, I_{n + 1})}{I_{n + 1}} = 0.
    \end{equation}

    Considering $p(\theta_n, I_{n + 1}) \equiv 0$ and $h(\theta_n, I_{n + 1}) = \sin(\theta_n)$, and changing $K(I_{n + 1})$ we obtain different systems well known in the literature, such as
    \begin{itemize}
        \item $K(I_{n + 1}) = I_{n + 1}$: the standard mapping~\cite{chirikovstdmap}.
        \item $K(I_{n + 1}) = 2/I_{n + 1}$: the static wall approximation of the Fermi-Ulam model~\cite{fermiulam1,fermiulam2}.
        \item $K(I_{n + 1}) = \delta I_{n + 1}$: the simplified bouncer model~\cite{bouncermodel1,bouncermodel2}.
        \item $K(I_{n + 1}) = I_{n + 1} + \delta I_{n + 1}^2$: the logistic twist map~\cite{HOWARD1995256}.
        \item $K(I_{n + 1}) = \abs*{I_{n + 1}}^{-\gamma}$: the Leonel mapping~\cite{Leonel2009,deOliveira2010,deOliveira2013,deOliveira2016,Leonel2020, Borin2023}.
    \end{itemize}

    In this paper, we consider the drift motion $\mathbf{E}\times\mathbf{B}$ in a toroidal plasma, where the electric field is generated by the combination of an external radial field $E$ and electrostatic potential $\tilde{\phi}$ generated by fluctuations in the plasma. The equations of the drift motion $\mathbf{E}\times\mathbf{B}$ are given by the following nontwist, area-preserving mapping~\cite{horton1998drift,10.1063/5.0147679,PhysRevE.109.015202}
    \begin{equation}
        \label{eq:mapa}
        \begin{aligned}
            I_{n + 1} &= I_n + \varepsilon\sin(2\pi\theta_n),\\
            \theta_{n + 1} &= \theta_n + \mu v(I_{n + 1})\left[\frac{M}{q(I_{n + 1})} - L\right] +\\
            &+\rho\frac{E(I_{n + 1})}{\sqrt{\abs{I_{n + 1}}}}\mod1,\\
        \end{aligned}
    \end{equation}
    where $q(I)$, $E(I)$, and $v(I)$ are nonmonotonic functions, representing the safety factor, external applied radial electric field, and toroidal plasma velocity, respectively. The non-monotonicity of these functions makes the system violate the twist condition, which implies different dynamical properties~\cite{Moser_1986,HOWARD1995256}. The nonmonotonic functions are given by,
    \begin{align*}   
        q(I) &= q_1 + q_2I^2 + q_3I^3,\\
        E(I) &= e_1I + e_2\sqrt{\abs{I}} + e_3,\\
        v(I) &= v_1 + v_2\tanh(v_3I + v_4),
    \end{align*}
    Although this mapping has many parameters, our only control parameter is $\varepsilon$ and the remaining parameters are chosen accordingly to Refs.~\cite{10.1063/5.0147679,PhysRevE.109.015202} and can be found in Table~\ref{tab:params}. The parameter $\varepsilon$ is proportional to the amplitude of the electrostatic instabilities that generate the drift motion. Thus, is the control parameter of integrability of the system. For the integrable case without perturbation ($\varepsilon = 0$), the variable $I$ is positive. However, with the perturbation $I$ can reach small negative values.

    \begin{table}[tb]
        \centering
        \caption{Parameter values of the two-dimensional area-preserving nontwist Hamiltonian mapping, Eq.~\eqref{eq:mapa}.}
        \label{tab:params}
        \begin{ruledtabular}
            \begin{tabular}{cccccc}
                $q_1$ & $5.0$ & $e_3$ & $4.13$ & $M$ & $15$ \\
                $q_2$ & $-6.3$ & $v_1$ & $-9.867$ & $L$ & $6$ \\
                $q_3$ & $6.3$ & $v_2$ & $17.47$ & $\mu$ & $1.83\times10^{-2}$ \\
                $e_1$ & $10.7$ & $v_3$ & $10.1$& $\rho$ & $-9.16\times10^{-1}$ \\
                $e_2$ & $-15.8$ & $v_4$ & $-9.0$ & --- & --- \\
            \end{tabular}
        \end{ruledtabular}
    \end{table}
    
    Similarly to the Leonel mapping~\cite{Leonel2009,deOliveira2010,deOliveira2013,deOliveira2016,Leonel2020, Borin2023}, when the action $I_n$ is small, the angles $\theta_n$ and $\theta_{n + 1}$ becomes uncorrelated due to the divergence of the second equation in Eq.~\eqref{eq:mapa}, and thus producing chaotic regions for nonzero perturbation. For larger values of the action, the angles become correlated and regularity appears as stability islands and invariant spanning curves. In Fig.~\ref{fig:phasespace} we show the phase space of the mapping~\eqref{eq:mapa} for different perturbation values. For $\varepsilon = 0$ (not shown) the mapping reduces to a nontwist radial map due to the nonmonotonicity of the functions $q(I)$, $E(I)$, and $v(I)$. In this case, the mapping is regular, the system is integrable, $I_{n}$ is constant, and the phase space has only periodic and quasiperiodic structures. On the other hand, for $\varepsilon > 0$ the regularity is broken, even for $\varepsilon \ll 1$, and the phase space becomes mixed. There is a coexistence of chaotic and regular domains and as $\varepsilon$ increases, the chaotic region expands in the vertical direction due to the breaking of the KAM curves. For the chosen parameter values, the phase space is bounded, \textit{i.e.}, and the transport in the vertical direction is confined to a finite area. Interestingly, the limits of the chaotic component shown in Fig.~\ref{fig:phasespace} are not symmetric with respect to the $I = 0$ line.
    
    The majority of stability islands for the considered values of $\varepsilon$ correspond to period-1 islands, which are centered around fixed elliptic points. The fixed points can be found from the following conditions:
    \begin{equation}
        \label{eq:fixedpoints}
        \begin{aligned}
            I_{n + 1} &= I_n = I^*,\\
            \theta_{n + 1} &= \theta_n = \theta^* + m,
        \end{aligned}
    \end{equation}
    where $m$ is an integer. Substituting Eqs.~\eqref{eq:fixedpoints} into Eqs.~\eqref{eq:mapa}, the fixed points must satisfy
    \begin{equation}
        \label{eq:fixedpoints2}
        \begin{aligned}
            \sin(2\pi\theta^*) &= 0,\\
            \mu v(I^*)\left[\frac{M}{q(I^*)} - L\right] + \rho\frac{E(I^*)}{\sqrt{\abs{I^*}}} &= m.
        \end{aligned}
    \end{equation}
    The first equation can be easily solved to find $\theta^* = 0, 0.5$. The second equation, however, cannot be solved analytically. We solve it numerically using the \emph{fsolve} function from the \emph{Scipy} module~\cite{scipy}, which uses Powell's hybrid method, and the red dots in Fig.~\ref{fig:phasespace} correspond to the elliptical fixed points found for different integers $m$.

    \begin{figure}[tb]
        \centering
        \includegraphics[width=\linewidth]{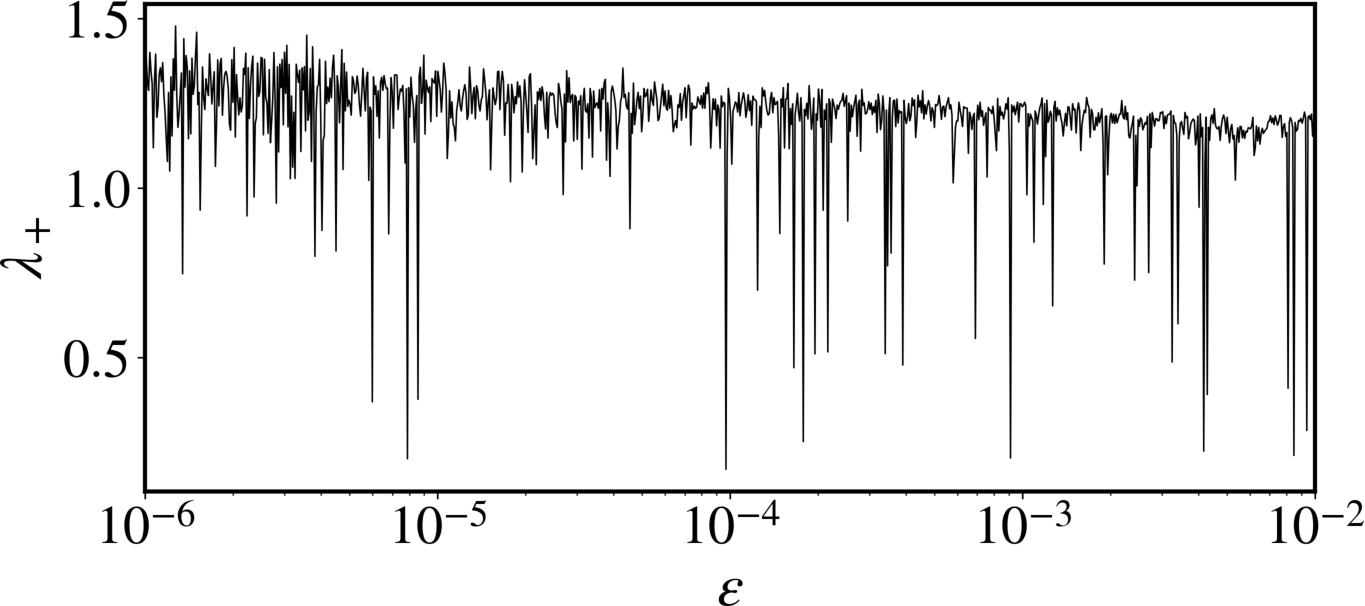}
        \caption{Largest Lyapunov exponent as a function of $\varepsilon$ with initial condition $(\theta_0, I_0) = (0.5, 1.0\times10^{-10})$ and total iteration time $N = 1.0\times10^{8}$.}
        \label{fig:lyapunovvseps}
    \end{figure}

    We characterize the chaotic component of the phase space using the Lyapunov exponents~\cite{Shimada1979, Benettin1980,Wolf1985,Eckmann1985}. Given a mapping $\vb{f}:\mathbb{R}^d \rightarrow \mathbb{R}^d$, defined as $\vb{x}_{n + 1} = \vb{f}(\vb{x}_n) = \vb{f}^n(\vb{x}_0)$, let $\vb{Df}^n$ be the $n$ iterate of the Jacobian matrix. We define the infinite-time Lyapunov exponents as
    \begin{equation}
        \lambda_i^{\infty} = \lim_{n\rightarrow\infty}\frac{1}{n}\ln\qty(\norm*{\vb{Df}^nu_i}),
    \end{equation}
    where $u_i$ is the eigenvector corresponding to the $i$th eigenvalue of $\vb{Df}^n$. A $d$-dimensional system has $d$ characteristic Lyapunov exponent, and we say the system is chaotic if at least one of them is positive. Furthermore, for Hamiltonian systems, which preserve volume in phase space under time evolution (Liouville's theorem)~\cite{lichtenberg2013regular}, the sum of all Lyapunov exponents must equal zero. Hence, for our two-dimensional area-preserving mapping, there are two characteristic exponents and they satisfy $\lambda_+ \equiv \lambda_1 = -\lambda_2$. In this case, all regular orbits, \textit{i.e.}, periodic, or quasi-periodic have zero Lyapunov exponents for infinite times, while chaotic orbits exhibit $\lambda_+ > 0$. In Fig.~\ref{fig:lyapunovvseps} we show the largest Lyapunov exponent, $\lambda_+$, as a function of the control parameter $\varepsilon$ for a single initial condition. We change $\varepsilon$ in a large interval and notice that $\lambda_+$, on the other hand, does not change significantly when compared to the range of variation of $\varepsilon$. This nearly constant value of $\lambda_+$ indicates that the chaotic component in phase space is scaling invariant with respect to the control parameter $\varepsilon$~\cite{Leonel2020}. We study such scaling properties in Sections~\ref{sec:survprob} and~\ref{sec:Irms}.

    \section{Survival probability and ratchet current}
    \label{sec:survprob}
  
\begin{figure}[tb]
        \centering
        \includegraphics[width=\linewidth]{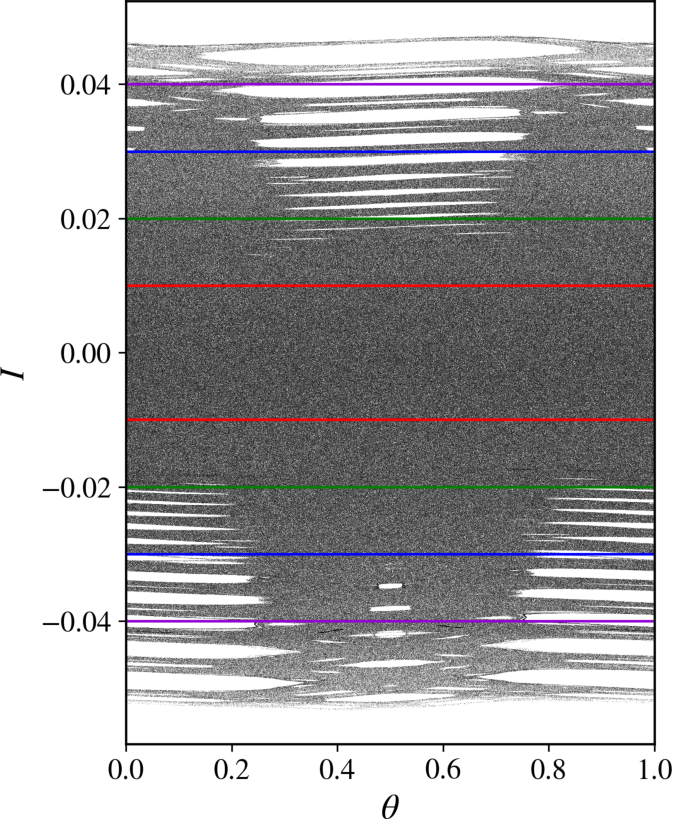}
        \caption{The phase space of the mapping Eq.~\eqref{eq:mapa} for 375 randomly chosen initial conditions within the chaotic sea with $\varepsilon = 1.0\times10^{-3}$. The regions delimited by the colored horizontal lines correspond to the survival regions defined by $(\theta, I) \in [0, 1) \times [-I_{\mathrm{esc}}, I_{\mathrm{esc}}]$ with (red) $I_{\mathrm{esc}} = 0.01$, (green) $I_{\mathrm{esc}} = 0.02$, (blue) $I_{\mathrm{esc}} = 0.03$, and (violet) $I_{\mathrm{esc}} = 0.04$.}
        \label{fig:phasespaceLs}
    \end{figure}

    In this section, we explore the transport of particles in the chaotic component of phase space. We have seen in the previous Section that the phase space for $\varepsilon \ll 1$ is bounded, and therefore the particle is confined within a finite area. We consider two exits placed symmetrically from the $I = 0$ line, and they define the survival region, $(\theta, I) \in [0, 1) \times [-I_{\mathrm{esc}}, I_{\mathrm{esc}}]$ (Fig.~\ref{fig:phasespaceLs}). We consider an ensemble of $M = 10^6$ particles randomly distributed in $(\theta, I) \in [0, 1) \times [-1.0\times10^{-10}, 1.0\times10^{-10}]$. We evolve each particle to at most $N = 10^6$ times, and if the particle reaches one of the exits, \textit{i.e.}, if $I_n = \pm I_{\mathrm{esc}}$, it escapes and we interrupt the evolution of this particle and initialize another particle. We repeat this procedure until the whole ensemble is exhausted. From this ensemble, we calculate the survival probability, $P(n)$, that corresponds to the probability of a particle surviving along the dynamics in a given chaotic domain without escaping. In other words, it corresponds to the fraction of particles that have not yet reached one of the exits until the $n$th iteration. Mathematically, it is defined as $P(n) = N_{\mathrm{surv}}(n)/M$, where $N_{\mathrm{surv}}(n)$ is the number of particles that have survived until the $n$th iteration. The behavior of the survival probability depends strongly on the characteristics of phase space. For fully chaotic systems, the survival probability follows an exponential decay~\cite{LEONEL20121669,Livorati2014,Mendez-Bermudez_2015,LIVORATI2018225} given by
    \begin{equation}
        \label{eq:survprob2}
        P(n) = P_0 \exp(-\kappa n),
    \end{equation}
    where $\kappa$ is the decay rate. However, for systems with mixed phase space, the decay is slower and usually characterized by the emergence of a power law tail for long times~\cite{Cristadoro2008, Altmann2009,Avetisov2010, Livorati2012,Borin2023} or by stretched exponential~\cite{DETTMANN2012403,deFaria2016,LIVORATI2018225}. As has been discussed previously, the stickiness effect, due to the presence of stability islands embedded in the chaotic sea, affects the statistical properties of the transport of particles in phase space. The chaotic orbits might be trapped near these stability islands, leading to long survival times, and causing the deviation from the exponential decay.
    
    \begin{figure*}[t]
      \centering
      \includegraphics[width=\linewidth]{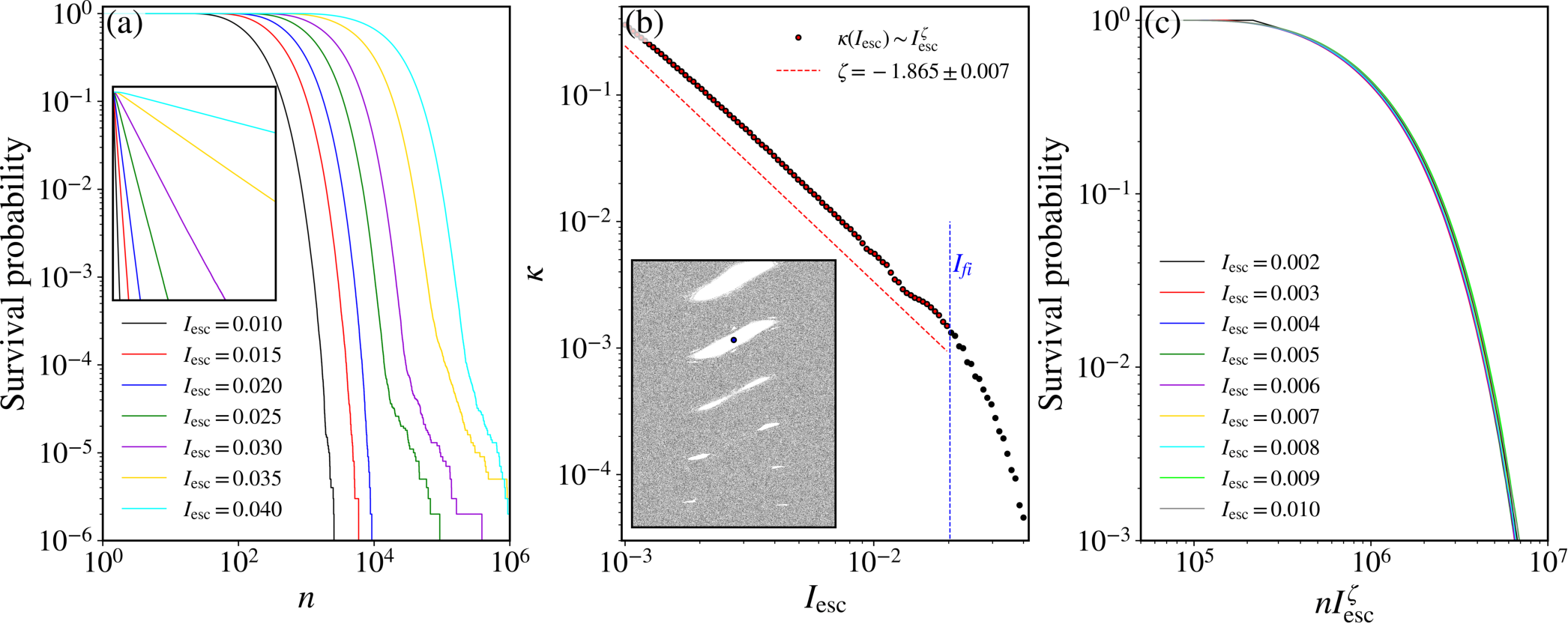}
      \caption{(a) The survival probability for different survival regions with $\varepsilon = 1.0\times10^{-3}$ considering an ensemble of $1.0\times10^6$ randomly chosen initial conditions on the interval $(\theta, I) \in [0, 1)\times[-1.0\times10^{-10}, 1.0\times10^{-10}]$. The inset corresponds to a semilog magnification of the region $(n, P) \in [0, 3 \times 10^4] \times [1.0 \times 10^{-3}, 1]$ to observe the exponential deay. (b) The escape rate, $\kappa$, is obtained from the optimal fitting considering the function $P(n) \sim e^{-\kappa n}$ for different values of $I_{\mathrm{esc}}$. For $I_{\mathrm{esc}}$ small enough, $\kappa$ obeys a power law (red dots), $\kappa(I_{\mathrm{esc}}) \sim I_{\mathrm{esc}}^\zeta$, with $\zeta = -1.865\pm0.007$. (c) The overlap of the survival probability for small $I_{\mathrm{esc}}$ onto a single and universal plot after the transformation $n \rightarrow nI_{\mathrm{esc}}^\zeta$. The inset in (b) is a magnification of Fig.~\ref{fig:phasespace}(a) within the region $(\theta, I) \in [0, 1)\times[0.015, 0.0225]$. The blue dot corresponds to the obtained center of the first period-1 island with coordinates given by $(\theta_{fi}, I_{fi}) = (0.5, 0.02026073030486245)$. For $I_{\mathrm{esc}} > I_{fi}$, $\kappa$ deviates from the power law decay (black dots).}
      \label{fig:survprob}
    \end{figure*}

    We calculate the survival probability for different survival regions, such as the ones shown in Fig.~\ref{fig:phasespaceLs}, with $\varepsilon = 1.0 \times 10^{-3}$ [Fig.~\ref{fig:survprob}(a)]. We observe the exponential decay for $I_{\mathrm{esc}} < 0.02$. For larger survival regions, the survival probability starts with an exponential decay for small times, until the power law tail emerges for long times. We note that the decay remains exponential until $I_{\mathrm{esc}}$ reaches the first period-1 stability island [blue dot on the inset of Fig.~\ref{fig:survprob}(b)], where the coordinates for its elliptic fixed point are $(\theta_{fi}, I_{fi}) = (0.5, 0.02026073030486245)$. The smaller islands below the period-1 island shown in the inset do not statistically influence the survival probability. Furthermore, the decay rate, $\kappa$, obtained from the optimal fitting based on the function given by Eq.~\eqref{eq:survprob2}, scales with $I_{\mathrm{esc}}$ as a power law for $I_{\mathrm{esc}} < I_{fi}$ with exponent $\zeta = -1.865 \pm 0.007$ [red dots and red dashed line in Fig.~\ref{fig:survprob}(b)]. The knowledge of $\zeta$ allows us to rescale the horizontal axis through the transformation $n \rightarrow nI_{\mathrm{esc}}^\zeta$ causing the survival probability curves to overlap onto a single, and hence, universal curve [Fig.~\ref{fig:survprob}(c)]. This indicates that the survival probability is scaling invariant for survival regions composed of predominantly chaotic regions ($I_{\mathrm{esc}} < I_{fi}$).

    Essentially, when some observable of a dynamical system exhibits scaling invariance, its expected behavior remains consistent and robust regardless of scale, \textit{i.e.}, we can rescale the system conveniently such that after a reparametrization, the observable is scale independent and exhibits universal features~\cite{scalinglaws}. The scaling invariance of the survival probability, for instance, has been explored in a variety of systems, such as the Leonel mapping~\cite{Borin2023}, billiard systems~\cite{Hansen2016}, and, more recently, on fractional versions of the standard map~\cite{MendezBermudez2023, Borin2024}.

\begin{figure*}[t]
        \centering
        \includegraphics[width=\linewidth]{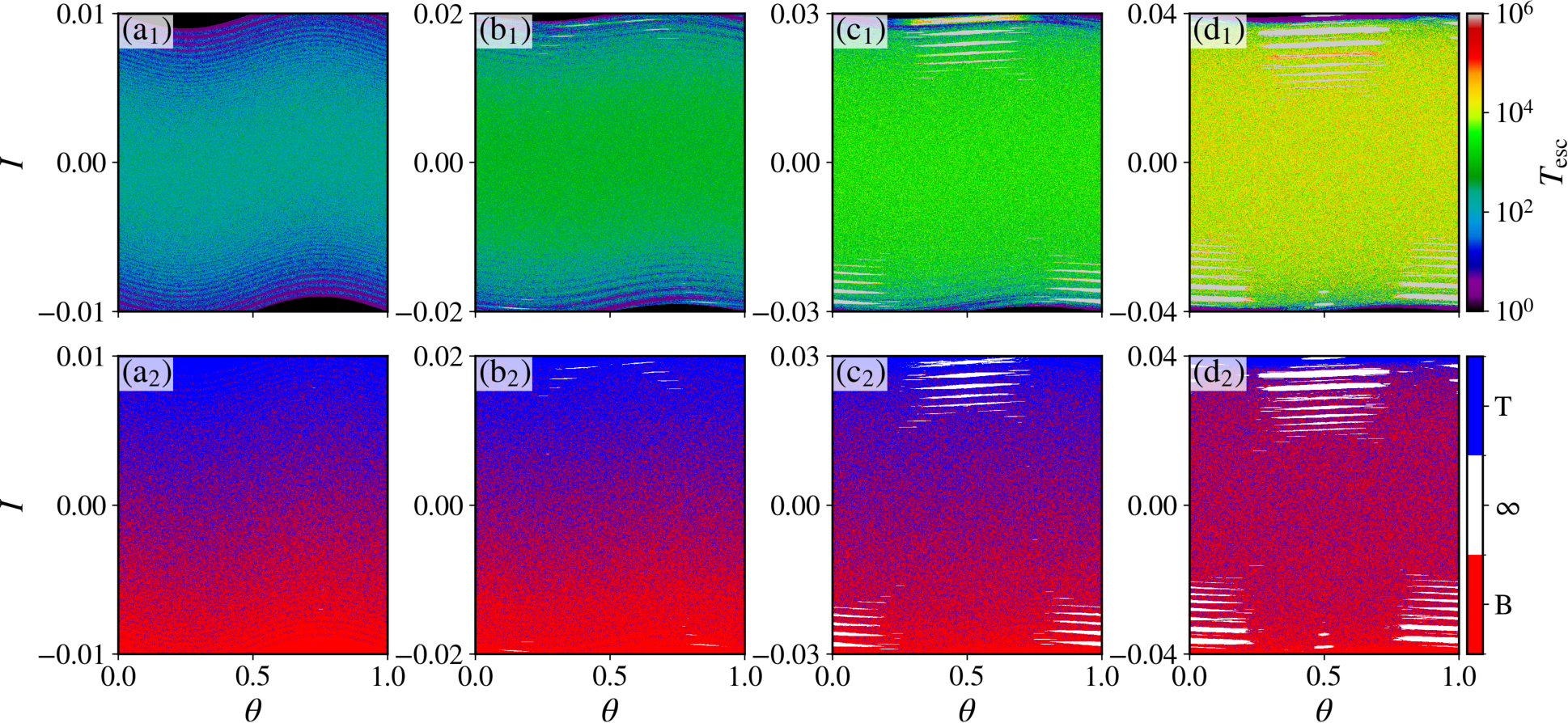}
        \caption{Escape times (top row) and escape basins (bottom) with $\varepsilon = 1.0\times10^{-3}$ for different survival regions, namely, (a) $I_{\mathrm{esc}} = 0.01$, (b) $I_{\mathrm{esc}} = 0.02$, (c) $I_{\mathrm{esc}} = 0.03$, and (d) $I_{\mathrm{esc}} = 0.04$. If the particle escapes through the bottom (top) exit, we color the point red (blue). If the particle never escapes the survival region, we color the initial condition white.}
        \label{fig:escape_times}
    \end{figure*}

    Let us now investigate how transport occurs for particles uniformly distributed on the survival regions $(\theta, I) \in [0, 1) \times [-I_{\mathrm{esc}}, I_{\mathrm{esc}}]$ for $I_{\mathrm{esc}} = 0.01, 0.02, 0.03, 0.04$. Note that $\langle I \rangle \approx 0$ at $n = 0$ for this ensemble. We consider $1080 \times 1080$ initial conditions and iterate each one for at most $N = 10^6$ times. We count the time each particle takes to reach either of the exits (top row of Fig.~\ref{fig:escape_times}). Black and blue colors indicate a fast escape while the gray color corresponds to particles that have not yet escaped within $N = 10^6$ iterations. The intermediate values of $T_{\mathrm{esc}}$ correspond to trapped particles. We observe that in cases of small survival regions, where stability islands are absent or nearly absent, the escape is fast. On the other hand, for the larger survival region considered [Fig.~\ref{fig:escape_times}(d$_1$)], most particles remain inside it for long times, with emphasis on the initial conditions near the stability islands (red color).

    We also construct the escape basin for these survival regions (bottom row of Fig.~\ref{fig:escape_times}). Particles that escape through the bottom (top) exit are colored red (blue), while those that never escape are colored white. Visually, for small survival regions [Figs.~\ref{fig:escape_times}(a$_2$) and~\ref{fig:escape_times}(b$_2$)] the red and blue points are distributed in equal measure, without a preferred region. However, for larger survival regions [Figs.~\ref{fig:escape_times}(c$_2$) and~\ref{fig:escape_times}(d$_2$)], with the emergence of the stability islands, the red basin appears to have a greater measure than the blue one, indicating a preference for escape through the bottom exit. In Table~\ref{tab:esc_side} we show the fraction of initial conditions that escape through the bottom ($p_\mathrm{B}$) and top ($p_\mathrm{T}$) exits, and also the ones that never escape ($p_\infty$), calculated from the escape basins in Fig.~\ref{fig:escape_times}. Note that $p_\mathrm{B} + p_\mathrm{T} + p_\infty = 1$. Indeed, the larger the survival region, \textit{i.e.}, the stronger the influence of the stability islands, the larger the ratio $p_\mathrm{B}/p_\mathrm{T}$, which corroborates our qualitative analysis of the escape basins: there is a preferential direction for the transport of particles in the chaotic component in phase space.

    \begin{table}[t]
        \centering
        \caption{Fraction of initial conditions that escape through the bottom ($p_\mathrm{B}$) and top ($p_\mathrm{T}$) exits and that never escapes ($p_\infty$) calculated from the escape basins in Fig.~\ref{fig:escape_times}.}
        \label{tab:esc_side}
        \begin{ruledtabular}
            \begin{tabular}{ccccc}
                Fig.~\ref{fig:escape_times} & $p_\mathrm{B}$ & $p_\mathrm{T}$ & $p_\infty$ & $p_\mathrm{B}$/$p_\mathrm{T}$ \\
                \midrule
                (a$_2$) & 0.498364 & 0.501601 & 0.000035 & 0.993548 \\
                (b$_2$) & 0.513745 & 0.484148 & 0.002107 & 1.061132 \\
                (c$_2$) & 0.502372 & 0.467406 & 0.030222 & 1.074810 \\
                (d$_2$) & 0.606776 & 0.327269 & 0.065954 & 1.854058 \\
            \end{tabular}
        \end{ruledtabular}
    \end{table}

    \begin{figure}[tb]
        \centering
        \includegraphics[width=\linewidth]{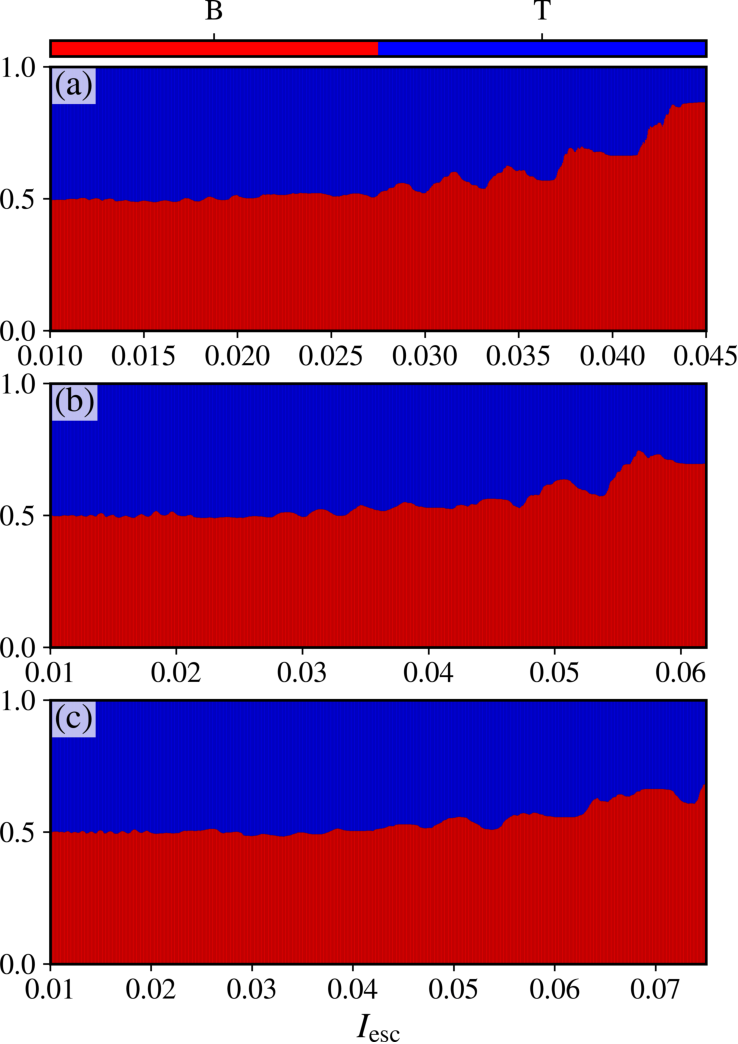}
        \caption{Fraction of initial conditions that escape through the (red) bottom exit and the (blue) top exit as a function of the exit's position with (a) $\varepsilon = 1.0\times10^{-3}$, (b) $\varepsilon = 2.0\times10^{-3}$, and (c) $\varepsilon = 3.0\times10^{-3}$. We considered an ensemble of $10^6$ randomly chosen initial conditions on the interval $(\theta, I) \in [0, 1]\times[-1.0\times10^{-10}, 1.0\times10^{-10}]$.}
        \label{fig:escape_side}
    \end{figure}

    However, how the particles are distributed in phase space might influence our previous conclusion. For example, some particles could be closer or farther away from the exits, while others might remain trapped for a longer duration. To eliminate this potential bias, we consider an ensemble of $10^6$ randomly chosen initial conditions on the interval $(\theta, I) \in [0, 1] \times [-1.0\times10^{-10}, 1.0\times10^{-10}]$, with $\langle I \rangle \approx 0$. Each particle is iterated up to $N = 10^6$ times. We compute the fraction of particles that escape through either the bottom or top exit as a function of the survival region limits, $I_{\mathrm{esc}}$, for $\varepsilon = 1.0 \times 10^{-3}$, $\varepsilon = 2.0 \times 10^{-3}$, and $\varepsilon = 3.0 \times 10^{-3}$ (Fig.~\ref{fig:escape_side}). For small survival regions, we observe that the escapes are evenly distributed between both the bottom and top exits. As $I_{\mathrm{esc}}$ increases, there is a tendency for the particles to escape through the bottom exit, more prominently for $\varepsilon = 1.0 \times10^{-3}$ [Fig.~\ref{fig:escape_side}(a)], characterizing the ratchet effect.
    
    \begin{figure}[t]
        \centering
        \includegraphics[width=\linewidth]{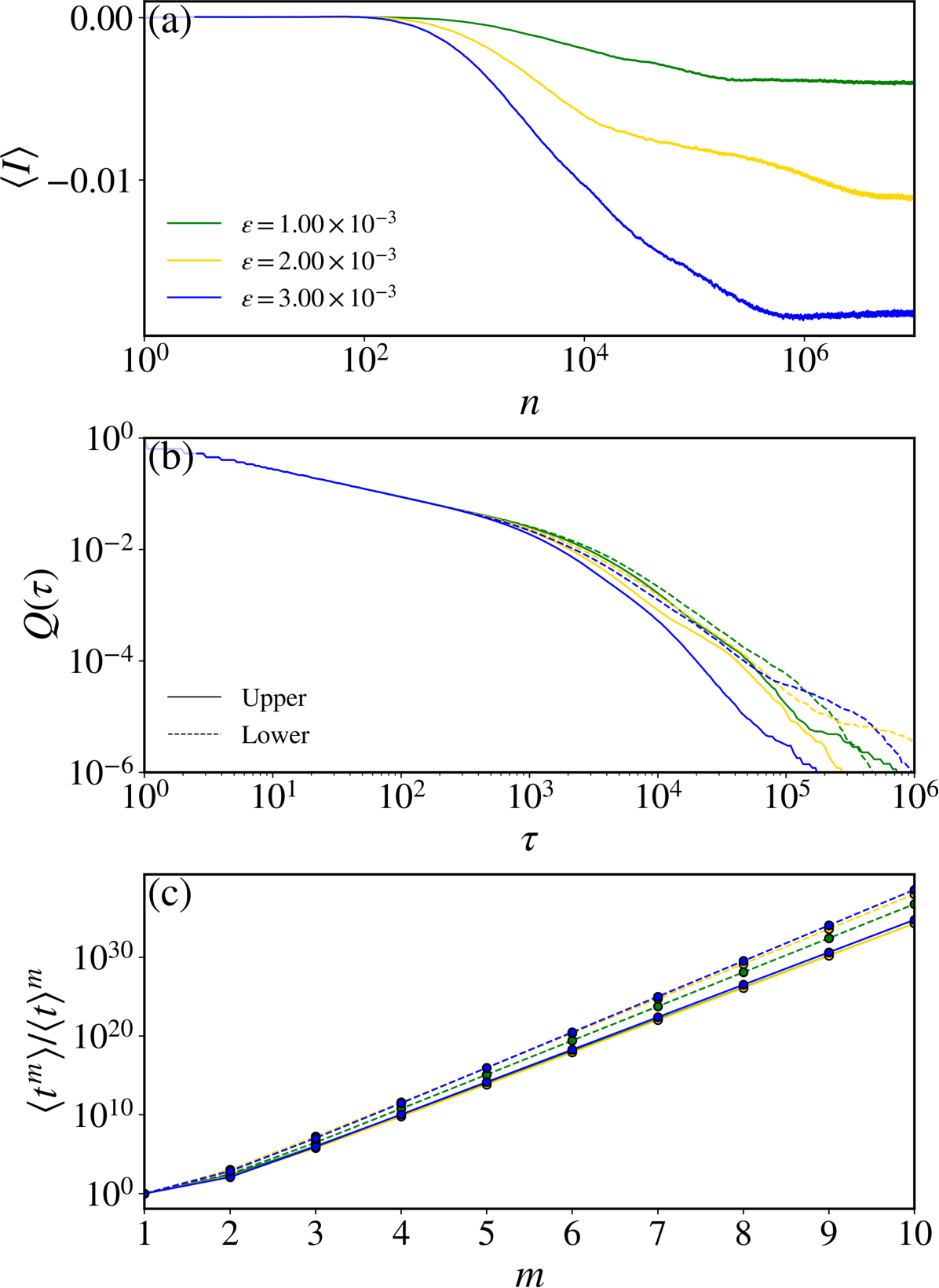}
        \caption{(a) The average of the action for an ensemble of $M = 10^6$ initial conditions randomly distributed on $I = 1.0\times10^{-10}$ at $n = 0$ for (green) $\varepsilon = 1.0\times10^{-3}$, (yellow) $\varepsilon = 2.0\times10^{-3}$, and (blue) $\varepsilon = 3.0\times10^{-3}$. (b) The cumulative distribution of recurrence times for the (full lines) top ($I > 0$) and (dashed lines) bottom ($I < 0$) regions of phase space. (c) Moments of the distribution of recurrence times normalized to $\langle t\rangle$.}
        \label{fig:rectimes}
    \end{figure}
    
    According to Gong and Brumer~\cite{Gong2004}, the average of $I$ over an ensemble of initial conditions, $\langle I \rangle$, represents the net current and the ratchet effect can also be characterized by a nonzero mean action. Thus, we calculate the average of the action as a function of time for the same previously used ensemble of $10^6$ randomly chosen initial conditions for the same three distinct values of $\varepsilon$ used in Fig.~\ref{fig:escape_side} [Fig.~\ref{fig:rectimes}(a)]. We observe a nonzero net current for all three cases for times $n > 10^2$. The mean value of $I$ increases with $\varepsilon$ because the volume of the chaotic component is larger for larger $\varepsilon$.

    Therefore, we have enough numerical evidence to state that the Hamiltonian nontwist mapping, given by Eq.~\eqref{eq:mapa}, exhibits the Hamiltonian ratchet effect. To understand the cause of this directed transport, we consider a single chaotic initial condition, $(\theta, I) = (0.5, 1.0 \times 10^{-10})$, and iterate it for $N = 10^9$ times with $\varepsilon = 1.0\times10^{-3}$, $\varepsilon = 2.0\times10^{-3}$, and $\varepsilon = 3.0\times10^{-3}$. We count the time it spends in both upper ($I > 0$) and lower ($I < 0$) regions of phase space. From this, we obtain two collections of recurrence times $\{t_j^{(\mathrm{U})}\}_{j = 1, 2, \ldots, N_t^{(\mathrm{U})}}$ and $\{t_j^{(\mathrm{L})}\}_{j = 1, 2, \ldots, N_t^{(\mathrm{L})}}$, where $N_t^{(\mathrm{U})}$ and $N_t^{(\mathrm{L})}$ are the number of recurrence times for the upper and lower regions, respectively \footnote{For the mentioned number of iterations, we obtained $N_t \sim 10^6$ for all values of $\varepsilon$.}, and define the probability distribution of recurrence times as $P^{(\textrm{U,L})}(t)$ for both collections of recurrence times. Alternatively, we define the cumulative distribution of recurrence times, $Q^{(\textrm{U,L})}(\tau)$, as follows:
    \begin{equation}
        Q^{(\textrm{U,L})}(\tau) = \sum_{t > \tau}P^{(\textrm{U,L})}(t) = \frac{N_{\tau}^{(\textrm{U,L})}}{N_t^{(\textrm{U,L})}},
    \end{equation}
    where $N^{(\textrm{U,L})}_\tau$ is the number of recurrence times larger than $\tau$, \textit{i.e.}, $t > \tau$ for the respective regions. Figure~\ref{fig:rectimes}(b) shows the cumulative distribution of recurrence times for the three values of $\varepsilon$ mentioned above for both collections of recurrence times, where the full lines correspond to the upper regions while the dashed lines to the lower one. We observe a power law tail for larger $\tau$, characteristic of systems that exhibit the stickiness effect~\cite{Afraimovich1997, Altmann2005, Altmann2006, Venegeroles2009, Abud2013, Lozej2020}. Furthermore, we observe different distributions for the upper and lower recurrence regions [full and dashed lines, respectively, in Fig.~\ref{fig:rectimes}(b)]: the decay is faster for the upper region. The difference between these two distributions becomes more evident when we calculate the higher moments of $P(t)$:
    \begin{equation}
        \langle t^m \rangle = \int_0^{t_{\mathrm{max}}}t^mP(t)\dd{t}.
    \end{equation}
    In Fig.~\ref{fig:rectimes}(c) we show the moments up to $m = 10$ normalized to $\langle t \rangle$. Starting from $m = 2$, the difference in the moments of the upper and lower distributions is already noticeable. Moreover, there is a substantial difference in the first moments as well (Table~\ref{tab:firstmoments}).

    \begin{table}[t]
        \centering
        \caption{The first moment, \textit{i.e.}, the mean recurrence time, of the recurrence time distributions for different values of $\varepsilon$ for the upper and lower regions of phase space.}
        \label{tab:firstmoments}
            \begin{tabular}{cccc}
                \toprule\toprule
                $\varepsilon$ & $\langle t \rangle^{(\mathrm{U})}$ & $\langle t \rangle^{(\mathrm{L})}$ & $\langle t \rangle^{(\mathrm{L})} - \langle t \rangle^{(\mathrm{U})}$ \\
                \midrule
                $1.0\times10^{-3}$ & $1.31 \times10^2$ & $1.51\times10^2$ & $0.20\times10^2$ \\
                $2.0\times10^{-3}$ & $1.02 \times10^2$ & $1.33\times10^2$ & $0.31\times10^2$ \\
                $3.0\times10^{-3}$ & $0.83 \times10^2$ & $1.21\times10^2$ & $0.38\times10^2$ \\
                \bottomrule\bottomrule
            \end{tabular}
    \end{table}

    Therefore, the difference in the upper and lower distribution is evidence of an unbalanced stickiness in phase space. This phenomenon makes chaotic orbits to sticky in the upper and lower region of phase space unevenly due to the spatial asymmetry with respect to the $I = 0$ line, creating the directed transport. Similar results have been reported in Ref.~\cite{Mugnaine2020} for the extended standard nontwist map. It is important to emphasize, however, that is the existence of asymmetric islands that leads to such a phenomenon, \textit{i.e.}, there is no symmetry transformation between the upper and lower islands. Additionally, the perturbation $\varepsilon$ and the survival region also influence the transport (Fig. \ref{fig:escape_side}). In regions without stability islands, the particles escape evenly from the bottom and top exits. Thus, the inherent asymmetry by itself is not enough to generate the ratchet current in our model.

    \section{Critical exponents and scaling law}
    \label{sec:Irms}

    In this section, we analyze the diffusion of chaotic orbits in phase space and the scaling properties of the chaotic component of phase space. We use the square root of the averaged squared action, defined as
    \begin{equation}
        \label{eq:Irms}
        I_{\mathrm{rms}} = \sqrt{\frac{1}{M}\sum_{i = 1}^M\frac{1}{n}\sum_{j = 1}^nI_{i, j}^2},
    \end{equation}
    where $M$ corresponds to an ensemble of initial conditions and $n$ is the length of the time series, as our observable.
    
    \begin{figure}[t]
        \centering
        \includegraphics[width=\linewidth]{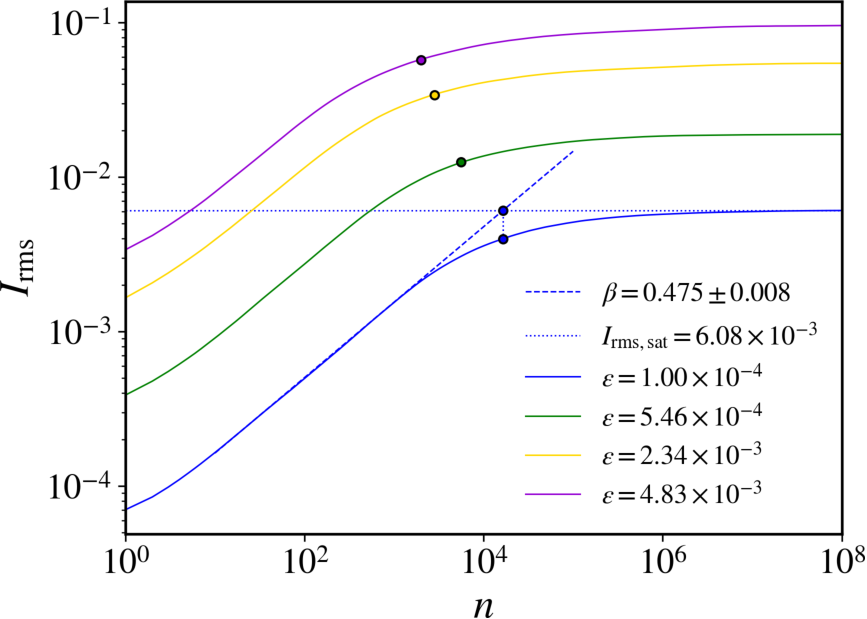}
        \caption{The $I_{\mathrm{rms}}$ as a function of the number of iterations for four distinct values of the perturbation $\varepsilon$. The colored dots indicate the transition point from the growth regime to the constant plateau of saturation. These points are determined by the intersection of the line obtained from the fitting of $I_{\mathrm{rms}}$ versus $n$ in the growth regime and the horizontal line $I_{\mathrm{rms}} = I_{\mathrm{rms,sat}}$.}
        \label{fig:Irms}
    \end{figure}

    The behavior of $I_{\mathrm{rms}}$ is shown in Fig.~\ref{fig:Irms} for different perturbation values $\varepsilon$. The curves exhibit an accelerated growth regime for small $n$ whereas for large $n$, they achieve a saturation limit, characterized by a constant plateau due to the bounded phase space area. The transition from accelerated growth to the constant plateau is given by a characteristic crossover number $n_x$. The behavior discussed above can be characterized by four critical exponents, namely, $\alpha$, $\beta$, $\gamma$, and $z$. To obtain them, we assume the following scaling hypotheses~\cite{Leonel2004,Leonel2004b,Leonel2007, Livorati2008, deOliveira2010, Leonel2015, Miranda2022}
    \begin{enumerate}
        \item For times $n \ll n_x$, $I_{\mathrm{rms}}$ scales as
        \begin{equation}
            \label{eq:1stscalinghyp}
            I_{\mathrm{rms}} \sim (n\varepsilon^\alpha)^\beta, 
        \end{equation}
        where $\alpha$ is the transport exponent and $\beta$ corresponds to the acceleration exponent.
        \item For $n \gg n_x$ the curve saturates, and the saturation value depends on $\varepsilon$ as
        \begin{equation}
            \label{eq:2ndscalinghyp}
            I_{\mathrm{rms,sat}} \sim \varepsilon^\gamma,
        \end{equation}
        where $\gamma$ is the saturation exponent.
        \item Finally, there is a changeover from an accelerated regime of growth to a constant plateau, identified by the crossover number $n_x$ and
        \begin{equation}
            \label{eq:3rdscalinghyp}
            n_x \sim \varepsilon^z,
        \end{equation}
        where $z$ is the crossover exponent.
        
    \end{enumerate}

    Therefore, using these scaling hypotheses, we describe the behavior of $I_{\mathrm{rms}}$ by a generalized homogeneous function, given by \cite{Leonel2004,Leonel2004b,Leonel2007}
    \begin{equation}
        \label{eq:Irmsgen}
        I_{\mathrm{rms}}(n\varepsilon^\alpha, \varepsilon) = \ell I_{\mathrm{rms}}(\ell^an\varepsilon^\alpha, \ell^b\varepsilon),
    \end{equation}
    where $\ell$ is a scaling factor and $a$ and $b$ are characteristic exponents. It is convenient to set $\ell^an\varepsilon^\alpha = 1$, leading us to
    \begin{equation}
        \label{eq:ell1}
        \ell = (n\varepsilon^\alpha)^{-1/a}.
    \end{equation}
    By substituting Eq.~\eqref{eq:ell1} into Eq.~\eqref{eq:Irmsgen}, we obtain
    \begin{equation}
        \label{eq:Irmsgen2}
        I_{\mathrm{rms}}(n\varepsilon^\alpha, \varepsilon) = (n\varepsilon^\alpha)^{-1/a} I_{\mathrm{rms}}(1, (n\varepsilon^\alpha)^{-b/a}\varepsilon).
    \end{equation}
    We assume $I_{\mathrm{rms}}(1, (n\varepsilon^\alpha)^{-b/a}\varepsilon) = \mathrm{const}$ for $n \ll n_x$, and comparing Eq.~\eqref{eq:Irmsgen2} with the first scaling hypothesis, Eq.~\eqref{eq:1stscalinghyp}, we find $a = -1/\beta$.

    Analogously, we set $\ell^b\varepsilon = 1$ and obtain
    \begin{equation}
        \label{eq:ell2}
        \ell = \varepsilon^{-1/b}.
    \end{equation}
    Using Eq.~\eqref{eq:ell2}, we can rewrite Eq.~\eqref{eq:Irmsgen} as
    \begin{equation}
        \label{eq:Irmsgen3}
        I_{\mathrm{rms}}(n\varepsilon^\alpha, \varepsilon) = \varepsilon^{-1/b} I_{\mathrm{rms}}(\varepsilon^{-a/b}n\varepsilon^\alpha, 1),
    \end{equation}
    where $I_{\mathrm{rms}}(\varepsilon^{-a/b}n\varepsilon^\alpha, 1)$ is assumed to be constant for $n \gg n_x$ (saturation regime). Therefore, upon comparing Eqs.~\eqref{eq:Irmsgen3} and the second scaling hypothesis, Eq.~\eqref{eq:2ndscalinghyp}, we find $b = -1/\gamma$.

    Therefore, the critical exponent $z$ can be obtained by combining the two scaling factors, Eqs.~\eqref{eq:ell1} and~\eqref{eq:ell2}, together with the values of $a$ and $b$ we have just found, at the transition point $n = n_x$. Indeed, we obtain
    %
    %
    %
    \begin{equation}
        \label{eq:nx}
        n_x = \varepsilon^{\frac{\gamma}{\beta} - \alpha}.
    \end{equation}
    Finally, by comparing Eq.~\eqref{eq:nx} with the third scaling hypothesis, Eq.~\eqref{eq:3rdscalinghyp}, we obtain the following scaling law
    \begin{equation}
        \label{eq:scalinglaw}
        z = \frac{\gamma}{\beta} - \alpha.
    \end{equation}

        \begin{figure*}[tb]
        \centering
        \includegraphics[width=\linewidth]{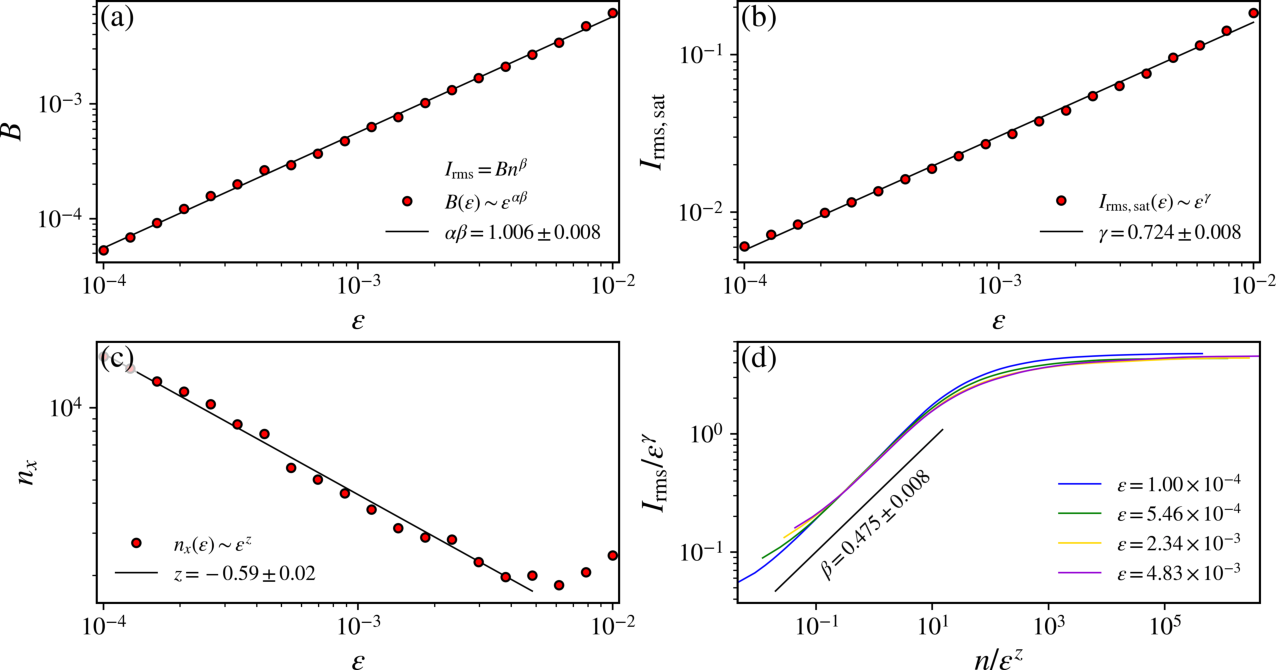}
        \caption{(a) The exponent $B(\varepsilon)$ from the power law fitting of $I_{\mathrm{rms}}$ versus $n$ (Fig.~\ref{fig:Irms}), $I_{\mathrm{rms}} = Bn^\beta$ as a function of $\varepsilon$. It scales with $\varepsilon$ as $B \sim \varepsilon^{\alpha\beta}$, and from the optimal fitting we obtain $\alpha\beta = 1.006\pm0.008$. (b) The saturation value of $I_{\mathrm{rms}}$, $I_{\mathrm{rms,sat}}$ as a function of $\varepsilon$. The saturation value is obtained for $n \gg n_x$ and it scales as a power law, with exponent $\gamma = 0.724 \pm 0.008$. (c) The transition point, $n_x$, from the growth regime to the constant plateau of saturation (colored dots in Fig.~\ref{fig:Irms}) as a function of $\varepsilon$. It depends on $\varepsilon$ as a power law, and from the optimal fitting we obtain for the crossover exponent $z = -0.59 \pm 0.02$. (d) The overlap of all curves in Fig.~\ref{fig:Irms} onto a single, and hence, universal curve after the transformations $n \rightarrow n/\varepsilon^z$ and $I_{\mathrm{rms}} \rightarrow I_{\mathrm{rms}}/\varepsilon^\gamma$.}
        \label{fig:SI}
    \end{figure*}

    Through numerical simulations, we obtain all these exponents. The acceleration exponent $\beta$ is obtained by plotting $I_\mathrm{rms}$ versus $n$ (Fig.~\ref{fig:Irms}) and performing a power law fitting $f(n) = Bn^A$ over the regime of growth. The exponent $\beta$ brings important information about diffusion. If $\beta = 0.5$, normal diffusion dominates the dynamics. Exponents larger (smaller) than 0.5 means super (sub) diffusion. We find that the exponent $\beta$ corresponds to $A$, and equals $\beta = 0.475 \pm 0.008$~\footnote{We performed the power law fitting for several values of $\varepsilon$ (those indicated in Fig.~\ref{fig:SI}) and this value corresponds to $\expval{\beta} \pm \sigma_\beta$, where $\expval{\cdot}$ is the mean and $\sigma_\beta$ is the standard deviation.}, slightly less than 0.5, meaning the diffusion process is sub diffusive. From the same power law fitting, we find $B = \varepsilon^{\alpha\beta}$. From different values of $\varepsilon$ we obtain different values of $B$ and from a power law fitting of $B$ versus $\varepsilon$ [Fig.~\ref{fig:SI}(a)], we obtain $\alpha\beta = 1.009\pm0.009$, such that the transport exponent is $\alpha = 2.12 \pm 0.05$.


    The saturation exponent, $\gamma$, is obtained using the second scaling hypothesis, Eq.~\eqref{eq:2ndscalinghyp}. We plot the saturation point of $I_{\mathrm{rms}}$, $I_{\mathrm{rms, sat}}$, as a function of $\varepsilon$ and perform a power law fitting [Fig.~\ref{fig:SI}(b)] to obtain $\gamma = 0.724 \pm 0.008$. As for the crossover exponent, $z$, defined by the third scaling hypothesis, Eq.~\eqref{eq:3rdscalinghyp}, we plot the point $n_x$ that marks the transition from a growth regime to the constant plateau of saturation observed in Fig.~\ref{fig:Irms} as a function of $\varepsilon$ [Fig.~\ref{fig:SI}(c)]. These points are indicated as colored dots in Fig.~\ref{fig:Irms}. Through a power law fitting, we obtain $z = -0.59\pm0.02$. Instead of performing the last mentioned power law fitting to find $z$, we could use the scaling law, given by Eq.~\eqref{eq:scalinglaw}, to obtain $z$ given $\alpha$, $\beta$, and $\gamma$. We obtain, in this case, $z = -0.6 \pm 0.1$, which remarkably agrees with the value obtained numerically, with a larger uncertainty, however. Furthermore, under the transformations $n\rightarrow n\varepsilon^z$, $I_{\mathrm{rms}} \rightarrow I_{\mathrm{rms}}/\varepsilon^\gamma$, the curves $I_{\mathrm{rms}}$ versus $n$ for different $\varepsilon$ overlap onto a single, and hence, universal curve [Fig.~\ref{fig:SI}(d)], confirming, therefore, the scaling invariance in the chaotic component of phase space.

    Let us now discuss our first scaling hypothesis, Eq.~\eqref{eq:1stscalinghyp}. We have assumed that $I_{\mathrm{esc}}$ scales with the perturbation $\varepsilon$ with an exponent of $\alpha\beta$. In several other works, the transport exponent has been assumed $\alpha = 2$ \textit{ad hoc} to validate the scaling assumptions~\cite{Leonel2004,Leonel2004b,Leonel2007, Livorati2008, deOliveira2010, Leonel2015, Miranda2022}, given that the acceleration exponent is $\beta = 0.5$, resulting in $\alpha\beta = 1$, for these cases. In a more recent work, Leonel \textit{et al.}~\cite{Leonel2020} demonstrated analytically the presence of the term $\varepsilon^2$ for the Leonel mapping. This leads to $\beta = 0.5$ in order to satisfy $\alpha\beta = 1$. Here, instead of assuming \textit{a priori} a value for $\alpha$ or for the product $\alpha\beta$, we have proposed a more general scaling hypothesis [Eq.~\eqref{eq:1stscalinghyp}] with which we have numerically obtained $\alpha\beta = 1$ (within numerical errors). In our case, the acceleration exponent slightly deviates from the value for the Leonel mapping, namely, $\beta = 0.475\pm0.008$, resulting in a transport exponent of $\alpha = 2.12 \pm 0.05$.

    \section{Conclusions}
    \label{sec:finalremarks}

    The $\mathbf{E}\times\mathbf{B}$ drift motion is observed in many plasma devices and plays a major role in the anomalous transport of particles. In magnetic confinement devices, $\mathbf{E}\times\mathbf{B}$ shear is responsible for reducing turbulence levels, improving the confinement. Thus, understanding this drift motion is crucial for improving plasma confinement and optimizing various applications. In this paper, we have explored the transport and diffusion of particles in the chaotic component in phase space in a newly reported nontwist Hamiltonian mapping, that describes the advected particle motion under the $\mathbf{E}\times\mathbf{B}$ drift. Firstly, our analysis focused on the properties of the survival probability considering two exits symmetrically placed on phase space with respect to the $I = 0$ line. We have demonstrated that the survival probability follows an exponential decay when the chaotic sea dominates the survival region. For survival region defined with $I_{\mathrm{esc}} > I_{fi}$, where $I_{fi}$ is the center of the elliptical fixed point of the first period-1 island, the survival probability follows a stretched exponential with a power law tail for long times. Furthermore, we have shown that the decay rate scales with $I_{\mathrm{esc}}$ as a power law for $I_{\mathrm{esc}} < I_{fi}$, with exponent $\zeta = -1.865 \pm 0.007$. By rescaling the horizontal axis of the plot $P(n) \times n$ by $n \rightarrow nI_{\mathrm{esc}}^\zeta$, we have found that the survival probability curves overlap into a single, and hence universal, plot for small survival regions, indicating that the survival probability maintains its behavior regardless of the size of the survival region.

    Secondly, we investigated whether the chaotic transport of particles has a preferential direction, \textit{i.e.}, whether the system exhibits unbiased transport. We have shown that considering particles distributed uniformly over the whole available phase space or particles randomly distributed over a small region around $I = 0$, both ensembles initially with $\langle I \rangle \approx 0$, the tendency is for the particles to escape through the bottom exit, thus exhibiting the so-called ratchet effect. The mechanics responsible for this effect is an unbalanced stickiness due to asymmetry of the chaotic component in phase space with respect to the line $I = 0$, \textit{i.e.}, the invariant spanning curves that bound the phase space are not symmetrical. This asymmetry generates a nonzero net current of particles due to different trapping times in the upper ($I > 0$) and lower ($I < 0$) regions of phase space. The cumulative distribution of recurrence times and the difference in the moments of the distribution of recurrence times support our claim for the existence of an unbalanced stickiness in phase space. To the best of our knowledge, the existence of ratchet current in the $\vb{E} \times \vb{B}$ drift motion has not been reported in the literature. This fact has complex implications for the impurity transport, heating, and instabilities in the plasma, and future studies are necessary to better understand them.

    Lastly, we presented a phenomenological description of diffusion in the chaotic component of phase space. We have chosen as our observable the square root of the averaged square action, $I_{\mathrm{rms}}$, and upon assuming three scaling hypotheses, we have found that the behavior of $I_{\mathrm{rms}}$ is characterized by four critical exponents. One of the exponents, $\beta$, characterizes the diffusion process. We have obtained $\beta = 0.475 \pm 0.008$, which is slightly below 0.5, indicating that the diffusion process is subdiffusive. The subdiffusive nature of the system is related to the trappings that generate the stickiness effect a chaotic orbit experiences throughout its evolution \cite{Spizzo_2019}. We have also derived an analytical scaling law relating these four exponents and from extensive numerical simulations, we have obtained all of them and showed that they remarkably agree with the scaling law. Our scaling hypotheses were supported by the collapse of the $I_{\mathrm{rms}}$ curves onto a single, and hence universal, curve. Additionally, even though the obtained scaling law [Eq.~\eqref{eq:scalinglaw}] is similar to the Family-Vicsek scaling discussed in Ref.~\cite{Barabási_Stanley_1995}, it is not directly connected. Nonetheless, this similarity motivates us to explore whether there is a correlation length that reaches the size of the bounded phase space area in the saturation regime. We intend to investigate this in a forthcoming paper.

    \section*{Declaration of competing interest}
    
    The authors declare that they have no known competing financial interests or personal relationships that could have appeared to influence the work reported in this paper.

    \section*{Code availability}

    The source code to reproduce the results reported in this paper is freely available on the Zenodo archive \cite{zenodo} and in the GitHub repository \cite{github}.

    \section*{Acknowledgments}

    This work was supported by the Araucária Foundation, the Coordination of Superior Level Staff Improvement (CAPES), the National Council for Scientiﬁc and Technological Development (CNPq), under Grant Nos. 01318/2019-0, 301019/2019-3, 403120/2021-7, 309670/2023-3, and by the São Paulo Research Foundation, under Grant Nos. 2018/03211-6, 2019/14038-6, 2022/03612-6, 2023/08698-9, 2023/16146-6.
    

%

\end{document}